\documentclass[longauth]{aa}  
\usepackage{natbib}
\bibpunct{(}{)}{;}{a}{}{,}
\usepackage{graphicx}
\usepackage{txfonts}
\usepackage{lipsum}
\usepackage{lscape}
\usepackage{placeins}
\usepackage[colorlinks=true, linkcolor=blue, citecolor=blue, urlcolor=blue]{hyperref}
\usepackage{multirow}

\begin{document}

   \title{A XRISM view of the iron line complex in NGC 1068: Rethinking the prototypical Compton-thick AGN}

\author{
S. Bianchi\inst{\ref{aff1}}
\and B. Vander Meulen\inst{\ref{aff2},\ref{aff3}}
\and E. Bertola\inst{\ref{aff13}}
\and V. Braito\inst{\ref{aff4},\ref{aff5},\ref{aff6}}
\and A. Comastri\inst{\ref{aff7}}
\and P. Cond\`o\inst{\ref{aff8},\ref{aff9}}
\and M. Dadina\inst{\ref{aff7}}
\and R. Della Ceca\inst{\ref{aff4}}
\and A. De Rosa\inst{\ref{aff20}}
\and V. E. Gianolli\inst{\ref{aff10}}
\and M. Guainazzi\inst{\ref{aff2}}
\and K. Iwasawa\inst{\ref{aff25},\ref{aff26}}
\and E. Kammoun\inst{\ref{aff11}}
\and M. Laurenti\inst{\ref{aff8},\ref{aff12},\ref{aff19}}
\and A. Luminari\inst{\ref{aff12},\ref{aff20}}
\and A. Marinucci \inst{\ref{aff21}}
\and G. Matt\inst{\ref{aff1}}
\and G. Matzeu\inst{\ref{aff27}}
\and R. Middei\inst{\ref{aff12}}
\and G. Miniutti\inst{\ref{aff23}}
\and E. Nardini\inst{\ref{aff13}}
\and F. Nicastro\inst{\ref{aff12}}
\and F. Panessa \inst{\ref{aff20}}
\and P.-O. Petrucci\inst{\ref{aff14}}
\and E. Piconcelli\inst{\ref{aff12}}
\and C. Pinto\inst{\ref{aff24}}
\and G. Ponti\inst{\ref{aff15},\ref{aff16},\ref{aff17}}
\and R. Serafinelli\inst{\ref{aff22},\ref{aff12}}
\and P. Severgnini\inst{\ref{aff4}}
\and D. Tagliacozzo\inst{\ref{aff1}}
\and F. Tombesi\inst{\ref{aff8},\ref{aff12},\ref{aff19}}
\and A. Tortosa\inst{\ref{aff12}}
\and F. Ursini\inst{\ref{aff1}}
\and C. Vignali\inst{\ref{aff18},\ref{aff7}}
\and L. Zappacosta\inst{\ref{aff12}}
}
\institute{
Dipartimento di Matematica e Fisica, Universit\`a degli Studi Roma Tre, Via della Vasca Navale 84, I-00146, Roma, Italy\\
\email{stefano.bianchi@uniroma3.it}\label{aff1}
\and European Space Agency (ESA), European Space Research and Technology Centre (ESTEC), Keplerlaan 1, 2201 AZ Noordwijk, The Netherlands\label{aff2}
\and Sterrenkundig Observatorium, Universiteit Gent, Krijgslaan 281 S9, 9000 Gent, Belgium\label{aff3}
\and INAF Osservatorio Astrofisico di Arcetri, Largo E. Fermi 5, 50125 Firenze, Italy\label{aff13}
\and INAF Osservatorio Astronomico di Brera, Via Bianchi 46 I-23807 Merate (LC), Italy\label{aff4}
\and Department of Physics, Institute for Astrophysics and Computational Sciences, The Catholic University of America, Washington, DC 20064, USA\label{aff5}
\and Dipartimento di Fisica, Universit\`a di Trento, Via Sommarive 14, 38123 Trento, Italy\label{aff6}
\and INAF Osservatorio di Astrofisica e Scienza dello Spazio di Bologna (OAS), Via Gobetti 93/3, I-40129 Bologna, Italy\label{aff7}
\and Dipartimento di Fisica, Universit\`a degli Studi di Roma ``Tor Vergata'', Via della Ricerca Scientifica 1, I-00133 Roma, Italy\label{aff8}
\and Dipartimento di Fisica, ``Sapienza'' Universit\`a di Roma, Piazzale Aldo Moro 2, I-00185 Roma, Italy\label{aff9}
\and INAF Istituto di Astrofisica e Planetologia Spaziali, Via del Fosso del Cavaliere 100, I-00133 Roma, Italy\label{aff20}
\and Department of Physics and Astronomy, Clemson University, Kinard Lab of Physics, Clemson, SC 29634, USA\label{aff10}
\and Institut de Ci\`encies del Cosmos (ICCUB), Universitat de Barcelona (IEEC-UB), Mart\'i i Franqu\`es, 1, 08028 Barcelona, Spain \label{aff25}
\and ICREA, Pg Llu\'is Companys 23, 08010 Barcelona, Spain\label{aff26}
\and Cahill Center for Astronomy \& Astrophysics, California Institute of Technology, 1216 East California Boulevard, Pasadena, CA 91125, USA\label{aff11}
\and INAF Osservatorio Astronomico di Roma, Via Frascati 33, 00078 Monte Porzio Catone (RM), Italy\label{aff12}
\and ASI Agenzia Spaziale Italiana, Via del Politecnico snc, 00133 Roma, Italy\label{aff21}
\and European Space Agency (ESA), European Space Astronomy Centre (ESAC), 28691 Villanueva de la Cañ    ada, Madrid, Spain\label{aff27}
\and Centro de Astrobiolog\'ia (CAB), CSIC-INTA, Camino Bajo del Castillo s/n, 28692, Villanueva de la Cañada, Madrid, Spain\label{aff23}
\and Universit\'e Grenoble Alpes, CNRS, IPAG, 38000 Grenoble, France\label{aff14}
\and INAF IASF Palermo, Via U. La Malfa 153, I-90146 Palermo, Italy\label{aff24}
\and Osservatorio Astronomico di Brera (INAF), Via E. Bianchi 46, Merate, 23807, Italy\label{aff15}
\and Max-Planck-Institut f\"ur extraterrestrische Physik, Giessenbachstrasse, Garching, 85748, Germany\label{aff16}
\and Como Lake Center for Astrophysics (CLAP), DiSAT, Universit\`a degli Studi dell'Insubria, via Valleggio 11, 22100 Como, Italy\label{aff17}
\and Instituto de Estudios Astrof\'isicos, Facultad de Ingenier\'ia y Ciencias, Universidad Diego Portales, Avenida Ej\'ercito Libertador 441, Santiago, Chile\label{aff22}
\and INFN - Rome Tor Vergata, Via della Ricerca Scientifica 1, 00133 Rome, Italy\label{aff19}
\and Dipartimento di Fisica e Astronomia `Augusto Righi', Universit\`a degli Studi di Bologna, via Gobetti 93/2, I-40129 Bologna, Italy\label{aff18}
}
 
\abstract
{NGC~1068 has long served as the reference Compton-thick Seyfert~2 galaxy. Decades of X-ray observations have established the presence of both neutral reflection and ionized circumnuclear emission. However, the limited spectral capabilities of previous CCD and grating data in the Fe--K band have prevented a clean separation of these components, thus leaving large uncertainties on key properties, such as the optical depth, structure, and kinematics of the reprocessing and emitting regions, and on their connection to nuclear outflows and feedback.}
{We exploited high-resolution X-ray spectroscopy to disentangle the neutral and ionized iron emission in NGC~1068; constrained the physical properties of the cold reflector through line energies, widths, fluorescence ratios, and Compton-shoulder diagnostics; and investigated the origin, kinematics, and potential feedback role of the highly ionized \ion{Fe}{xxv} and \ion{Fe}{xxvi} emission.}
{We analyzed a XRISM/Resolve observation of NGC~1068, focusing on the Fe~K$\alpha$ and Fe~K$\beta$ fluorescent lines and on the \ion{Fe}{xxv} and \ion{Fe}{xxvi} emission complexes. Line centroid energies, intrinsic widths, flux ratios, and constraints on the Compton shoulder were derived through local spectral fitting, and compared with atomic calculations and theoretical predictions.}
{The centroid energies of the Fe~K$\alpha$ and Fe~K$\beta$ lines tightly constrain the emitting material to be neutral or near neutral. The observed Fe~K$\beta$/K$\alpha$ ratio, together with the stringent upper limit on the Compton shoulder ($\lesssim$8--11\% of the core flux), disfavor reflection dominated by a homogeneous classical Compton-thick medium, indicating that most of the neutral Fe~K$\alpha$ emission arises in optically thin or moderately Compton-thick gas.
The \ion{Fe}{xxv} and \ion{Fe}{xxvi} emission lines exhibit remarkably large velocity widths, of several thousand km~s$^{-1}$. These broad profiles closely resemble the integrated optical and infrared [\ion{O}{iii}] and [\ion{O}{iv}] lines associated with the large-scale biconical outflow, and are naturally interpreted as the X-ray signature of a more highly ionized, faster, and more spatially confined phase of the same outflow.}
{The iron-K emission of NGC~1068 reveals a stratified circumnuclear environment in which neutral and highly ionized components arise in physically distinct regions.
The neutral Fe~K fluorescence originates predominantly in optically thin or mildly Compton-thick material, despite the persistently Compton-thick line-of-sight obscuration, indicating a geometrically complex cold reprocessor. The highly ionized iron emission lines trace a fast component consistent with a bipolar outflow on parsec scales, whose large velocities and inferred energetics suggest that it may represent an efficient channel for feedback. If similar ionized outflows and reflection geometries are common in heavily obscured Seyferts, they may constitute an underappreciated aspect of feedback and circumnuclear structure in this class of sources.}


\authorrunning{Bianchi et al.}
\maketitle
\nolinenumbers

\section{Introduction}
Obscured active galactic nuclei (AGN) provide a privileged view of the circumnuclear environment of accreting supermassive black holes (BHs). When the line-of-sight column density exceeds the Compton-thick threshold ($N_{\rm H} \gtrsim 1.5 \times 10^{24}\,\mathrm{cm^{-2}}$), the direct X-ray continuum is strongly attenuated below $\sim$10~keV and no longer dominates the observed spectrum. In this high-contrast regime, fluorescent emission lines, Compton-scattered continua, and emission lines from ionized gas stand out more prominently against the diminished primary continuum. This allows the properties of the reprocessing gas, including its column density, ionization state, chemical composition, geometry, and kinematics, to be constrained from the reprocessed emission itself \citep[e.g.,][]{Matt2000,Bianchi2012,RamosAlmeida2017}. Nearby Compton-thick AGN therefore represent prime laboratories for mapping the structure of obscuration and reflection and for linking the reprocessing media to the multi-phase outflows observed on larger scales.

High-resolution microcalorimeter spectroscopy represents a major advance in the X-ray study of obscured AGN. By delivering an energy resolution of a few electronvolts across the entire Fe~K band without dispersing the photons, and thus retaining the full effective area of the telescope, it enables direct measurements of line centroids, intrinsic widths, and profile asymmetries. The X-Ray Imaging and Spectroscopy Mission \citep[XRISM;][]{Tashiro2025} is the first mission to provide these capabilities for bright nearby AGN, through its Resolve instrument \citep{Kelley2025,Ishisaki2025}. These characteristics make XRISM/Resolve particularly well suited to disentangling neutral and ionized iron emission and to accessing directly the velocity structure of the reprocessing and emitting gas in heavily obscured systems.

NGC~1068 is the prototypical obscured AGN and a cornerstone of the orientation-based unification framework \citep{Antonucci1993}. Its optical classification as a Seyfert~2 galaxy, together with the detection of broad permitted emission lines only in polarized light, provided early and compelling evidence for the presence of a luminous but heavily obscured nucleus \citep{Antonucci1985}. Owing to its proximity and brightness, NGC~1068 has been studied in exceptional detail across the electromagnetic spectrum, revealing a complex circumnuclear environment extending from subparsec to kiloparsec scales. Optical and infrared observations have mapped a prominent narrow-line region with a well-defined biconical morphology and ordered velocity field, commonly interpreted as a large-scale outflow (e.g., \citealp{crenshaw_resolved_2000,das_kinematics_2006,Marconcini2025}). At smaller scales, infrared interferometry and high-resolution radio observations have provided direct evidence for a compact geometrically thick dusty structure surrounding the nucleus, commonly associated with the obscuring torus, and have revealed a complex morphology suggestive of clumpy dust and polar elongation \citep[e.g.,][]{Jaffe2004,GarciBurillo2016,LopezGonzaga2016}.

In X-rays, NGC~1068 is a textbook example of a Compton-thick AGN. Early Ginga and ASCA data showed that the direct nuclear continuum is heavily suppressed and that the spectrum is dominated by reprocessed emission, characterized by a flat hard X-ray continuum and an unusually strong Fe~K emission-line complex, including both neutral and ionized components \citep{Koyama1989,Ueno1994,Iwasawa1997}. XMM-Newton later demonstrated that multiple reflecting components spanning different column densities and ionization states are required, pointing to a complex and stratified circumnuclear reprocessing medium \citep{Matt2004}.
At hard X-ray energies, observations with BeppoSAX, Suzaku, and more recently NuSTAR have contributed to building a scenario in which the Compton reflection hump and the neutral Fe~K$\alpha$ emission do not originate entirely from the same material, pointing to the presence of multiple reflecting regions with different column densities and spatial distributions \citep{Bauer2015}. In addition, NuSTAR revealed variability of the hard X-ray emission on timescales of months to years, interpreted as changes in the column density or covering factor of the obscuring material along the line of sight \citep{Marinucci2016,Zaino2020}, thus providing direct evidence that the circumnuclear reprocessor in NGC~1068 is not static, but dynamically evolving. In addition, X-ray polarimetry with the \textit{Imaging X-ray Polarimetry Explorer} (IXPE) has revealed significant 2--8~keV polarization in NGC~1068, with a polarization angle perpendicular to the parsec-scale radio axis, confirming that the observed emission is dominated by scattered radiation and providing direct geometrical constraints on the circumnuclear obscurer, similarly to the Circinus galaxy \citep{Marin2024}.

High angular resolution Chandra imaging showed that the soft X-ray emission is spatially extended over several hundred parsecs, with a clear correspondence to the ionization cones observed at optical wavelengths (e.g., \citealp{Young2001a,Ogle2003}). This spatial association strongly suggests a physical connection between the soft X-ray--emitting gas and the large-scale outflowing narrow-line region, as observed in other obscured AGN \citep{Bianchi2006}. From a spectroscopic point of view, high-resolution observations with Chandra gratings and XMM-Newton/RGS revealed a rich emission-line spectrum dominated by He- and H-like transitions of light elements together with prominent radiative recombination continua, unambiguously indicating that photoionization is the dominant emission mechanism, as commonly found in obscured AGN (e.g., \citealp{GuainazziBianchi2007}). Detailed modeling of these spectra showed that multiple ionization components are required, spanning a broad range of ionization parameters and column densities, with little to negligible contribution from collisionally ionized plasma \citep{Kinkhabwala2002,Ogle2003,Kallman2014,grafton-waters_photoionisation_2021}. In this context, radiation pressure compression (RPC) provides a physically motivated framework to describe the soft X-ray emission in NGC~1068 and in obscured AGN more generally, naturally producing a stratified ionization structure that successfully reproduces the observed emission measure distribution (e.g., \citealp{Bianchi2019}).

In this work we present the first high-resolution X-ray microcalorimeter spectrum of NGC~1068 obtained with XRISM/Resolve. These data provide an unprecedented view of the Fe~K band in this prototypical Compton-thick AGN, enabling a detailed characterization of the neutral and ionized iron emission and their kinematic properties in the context of the complex, multi-phase circumnuclear environment described above.

\section{Observations and data reduction}

NGC~1068 was observed with the XRISM observatory on 2025 February 08 for a total on-source exposure of 115~ks (ObsID 201116010; PI: S. Bianchi). Both the Resolve microcalorimeter spectrometer \citep{Ishisaki2025}, in ``PX\_NORMAL'' data mode,  and the Xtend CCD imager \citep{Noda25}, in full window mode, operated simultaneously during the observation. The target was placed at the nominal Resolve aimpoint. At the time of the observation the Resolve gate valve was closed, which restricted the energy band to $\gtrsim$1.7~keV. The data were reprocessed with the XRISM pipeline included in \texttt{HEASoft} 6.36, using the calibration database (CALDB) 20250915. We applied standard screening criteria and a filter of the geomagnetic cutoff rigidity (COR) with a threshold of 6. The resulting net exposure is 110~ks.

For Resolve the source spectrum was extracted from the full field of view ($\sim3\arcmin$ diameter), excluding pixel number 27, as recommended to avoid calibration uncertainties. This aperture fully encompasses the extended X-ray emission of NGC~1068, including the narrow-line region and circumnuclear reflection structures known from previous high-resolution imaging.  Large (L) type redistribution matrix files (RMFs) and ancillary response files (ARFs) were produced with the pipeline tools, including gain and line spread function corrections.  We selected only the high-resolution  (Hp) events, following the XRISM data analysis guidelines; these account for 94\% of all the events during this
observation. The background spectrum was generated using the stacked non-X-ray background (NXB) event file provided by the XRISM team.\footnote{\url{https://heasarc.gsfc.nasa.gov/docs/xrism/analysis/nxb/resolve_nxb_db.html}} We first generated a good time interval (GTI) file for the source and the NXB events, with the same filtering criteria. Taking into account that the level of background may depend on the COR, we  filtered  the NXB event file with the same COR as per the source extraction. The task \texttt{rslnxbgen} was then used to extract the NXB spectrum from this screened background event file,  considering   only Hp  events and excluding  pixel 27. The Resolve light curve shows no evidence for flux variability during the 110 ks exposure. Nonetheless, we split the observation into two equal time segments and fitted the resulting spectra separately. No significant changes were detected in flux or spectral shape within the statistical uncertainties, and the full dataset was used for the final analysis.  

All spectra were grouped using \texttt{ftgrouppha} with optimal
binning \citep{Kaastra2016}, with the additional requirement of a minimum signal-to-noise ratio of 3, and fitted using the \citet{Cash1979} statistics as implemented in \texttt{XSPEC} v12.15.1 \citep{Arnaud1996}. Unless otherwise stated, errors are quoted at the 1$\sigma$ confidence level (c.l.) for one interesting parameter, and upper limits at 99\% c.l. The redshift of NGC~1068 was fixed at $z=0.003793$ (\citealt{Huchra1999}, based on \ion{H}{i} 21 cm measurements compiled by \citealt{Bottinelli1990}), but we assume a luminosity distance $\mathrm{D_L} = 10.1$ Mpc (\citealt{Padovani2024}, see also \citealt{Markham2026}). 

\section{\label{sec:spectralanalysis}Spectral analysis}

\begin{figure*}[ht]
    \centering
        {\includegraphics[width=\textwidth]{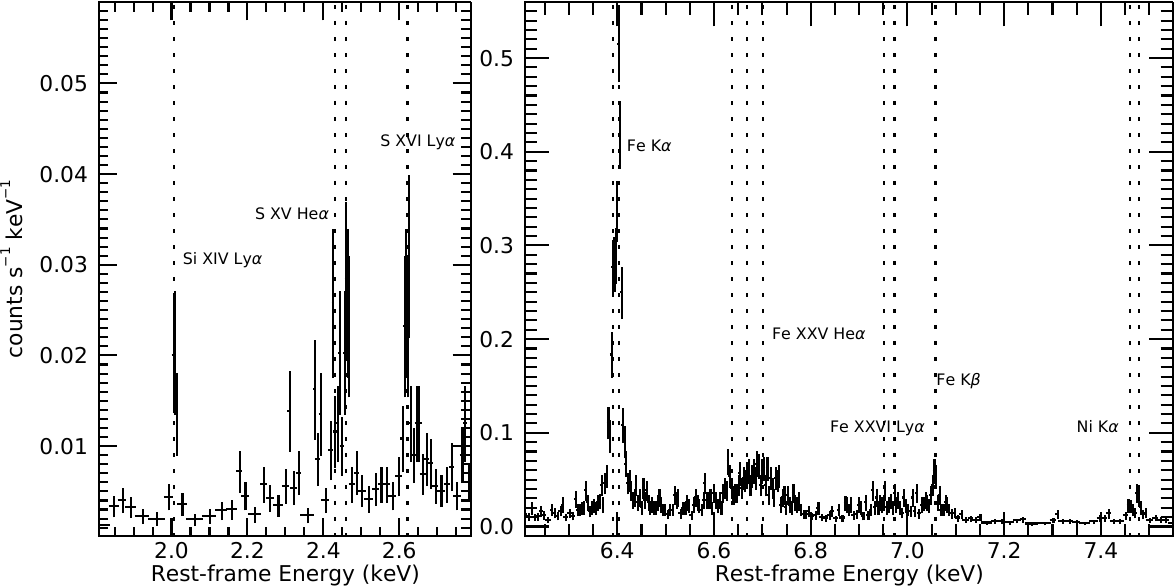}}
     \caption{XRISM/Resolve spectrum of NGC~1068 showing only emission lines detected at $\geq 99\%$ confidence in the blind line scan (Sect.~\ref{sec:spectralanalysis}). For clarity, the detected lines are displayed in two separate energy bands, which encompass all features meeting the detection criterion. Vertical dashed lines mark the centroid energies of the detected features, which are labeled in the figure (see Table~\ref{tab:lines}).}
      \label{fig1}
\end{figure*}

We performed a blind search for narrow emission and absorption features over the entire 1.7--12 keV Resolve band using a Gaussian line-scan procedure. At each trial energy, a narrow Gaussian component was added to the best-fit continuum model and its normalization was stepped logarithmically over positive and negative values, while recording the change in fit statistic. The resulting two-dimensional maps were used to identify statistically significant features.

Emission lines detected at more than 99\% c.l. are shown and labeled in Fig.~\ref{fig1}, and listed in Table~\ref{tab:lines}. In addition to these lines, several other features are present at lower significance throughout the full 1.7--12 keV band. A more comprehensive and dedicated line-scan analysis, including a detailed assessment of statistical significance and physical interpretation, will be presented in a forthcoming paper. To aid the interpretation of the following results, we note that the in-flight calibration of Resolve demonstrates an energy determination accuracy of $\lesssim$0.2~eV \citep{Porter2025}, corresponding to $\lesssim$10~km~s$^{-1}$ in the Fe~K band.

\begin{table}[ht]
\caption{Measured parameters of the main emission lines in the XRISM/Resolve spectrum of NGC~1068.}
\label{tab:lines}
\centering
\setlength{\tabcolsep}{3pt} 
\begin{tabular}{lccc}
\noalign{\vskip 2pt}\hline\noalign{\vskip 2pt}
Line & Velocity shift & $\sigma$ & Flux or ratio \\
     & (km\,s$^{-1}$) & (km\,s$^{-1}$) &
($10^{-5}$ ph\,cm$^{-2}$\,s$^{-1}$) \\
\noalign{\vskip 2pt}\hline\noalign{\vskip 2pt}
Fe K$\alpha$ &
\multirow{3}{*}{$-30 \pm 12$\tablefootmark{$\ddagger$}} &
\multirow{3}{*}{$155^{+18}_{-16}$\tablefootmark{$\ddagger$}} &
$5.8 \pm 0.2$ \\
Fe K$\beta$ &
 &
 &
$0.103 \pm 0.016$\tablefootmark{$\dagger$} \\
Ni K$\alpha$ &
 &
 &
$0.066 \pm 0.012$\tablefootmark{$\dagger$} \\
Fe K$\alpha$ (broad) &
$-700^{+250}_{-300}$ &
$2400 \pm 400$ &
$0.39 \pm 0.06$\tablefootmark{$\dagger$} \\
\noalign{\vskip 2pt}\hline\noalign{\vskip 2pt}
\ion{Fe}{xxv} He$\alpha$ &
\multirow{2}{*}{$-300^{+300}_{-500}$\tablefootmark{$\ddagger$}} &
\multirow{2}{*}{$2400 \pm 200$\tablefootmark{$\ddagger$}} &
$4.4^{+0.6}_{-0.5}$ \\
\ion{Fe}{xxvi} Ly$\alpha$ &
 &
 &
$1.5^{+0.6}_{-0.4}$ \\
\noalign{\vskip 2pt}\hline\noalign{\vskip 2pt}
\ion{Si}{xiv} Ly$\alpha$ &
\multirow{3}{*}{$-100 \pm 140$\tablefootmark{$\ddagger$}} &
\multirow{3}{*}{$710^{+150}_{-120}$\tablefootmark{$\ddagger$}} &
$7.1\pm1.5$ \\
\ion{S}{xv} He$\alpha$ &
 &
&
$3.2\pm0.9$ \\
\ion{S}{xvi} Ly$\alpha$ &
&
 &
$1.4\pm0.3$ \\

\noalign{\vskip 2pt}\hline\noalign{\vskip 2pt}
\end{tabular}
\tablefoot{The Fe K$\alpha$ and K$\beta$ lines are modeled with \texttt{zbfeklor} and \texttt{zbfekblor}. The Ni K$\alpha$ and Ly$\alpha$ lines are modeled as 2:1 Gaussian doublets. The He$\alpha$ lines are modeled as Gaussian triplets (four components in the case of \ion{Fe}{xxv}). In all cases, the reported fluxes refer to the total flux summed over all components (see text for further details).\\
\tablefoottext{$\dagger$}{Flux ratios relative to the Fe K$\alpha$ core flux.}
\tablefoottext{$\ddagger$}{Velocity shift and Gaussian width are tied among lines of the same element or line family.}}
\end{table}

\subsection{\label{neutrallines}The neutral fluorescence lines}

We started with a local fit in the energy range $\sim6.0-6.5$ keV, adopting a power law with $\Gamma=1$\footnote{We fix $\Gamma=1$ as a convenient approximation to the locally flat shape of a pure reflection spectrum. We note that the choice of the photon index does not affect the results on the line parameters in such a restricted energy range.}
 and free normalization as the baseline continuum. The neutral Fe K$\alpha$ line is modeled using a \texttt{zbfeklor} component, available in \texttt{XSPEC}, which provides a phenomenological description of its intrinsic profile as measured in laboratory experiments and represents the current standard in high-resolution X-ray spectroscopy. In this model, the line is parameterized using seven Lorentzian components, yielding the most detailed representation of the K$\alpha_1$–K$\alpha_2$ doublet, which corresponds to the two main fluorescent transitions with an expected 2:1 intensity ratio. The relative energies and intensities of the components are fixed to the laboratory measurements of \citet{Holzer1997}. Higher-order satellite lines (K$\alpha_{3,4}$) are not included in this model; their expected contribution is $\lesssim$0.8\% of the main Fe K$\alpha$ emission \citep{Shigeoka2004,Diamant2006}, and is therefore negligible at the signal-to-noise of the present data. The only free parameters are the normalization and a velocity dispersion $\sigma$ in km s$^{-1}$, which broadens the overall line profile, as well as a velocity shift\footnote{All velocity shifts quoted in this work are measured relative to the systemic redshift adopted above, which is based on \ion{H}{i} 21 cm measurements tracing the large-scale kinematics of the host galaxy.} introduced with the convolution model \texttt{vashift}. The best-fit parameters are $\sigma=180^{+30}_{-20}$ and $v=-36\pm13$ km s$^{-1}$. Previous \textit{Chandra} High Energy Transmission Grating (HETG) analyses reported Fe K$\alpha$ widths of order $\sigma \sim 700$ km s$^{-1}$ or larger \citep{Ogle2003,Kallman2014,Andonie2022}, but \citet{Liu2016} showed that first-order HEG measurements are likely overestimated and that the line is consistent with being unresolved when higher-order spectra are considered.
 Significant residuals are still present after this fit, particularly at the blue wing and the core of the line. We note that similar residuals are also apparent  in the \textit{Chandra}/HETG data \citep{Kallman2014}.

On the other hand, any Compton shoulder (CS), if present, appears to be very weak, given the absence of prominent residuals on the red wing of the line. To quantitatively test this, we built a simple top-hat profile extending from 6.24 to 6.40 keV, normalized to the unit area, broadened and shifted as the narrow core. This shape mimics the roughly flat energy distribution of first-order Compton down-scattered photons expected from reflection in cold optically thick matter \citep[see, e.g.,][]{Matt2002}, while allowing a direct measurement of the total shoulder flux as a fraction of the core. The CS modeled in this way is not statistically required by the data, with an upper limit of $8\%$ of the core flux. Replacing this CS component with a template based on the smoother profile that accounts for bound-electron scattering, as implemented in SKIRT \citep{VanderMeulen2023},\footnote{In SKIRT, the bound-electron momentum distribution is based on Compton profiles from \cite{Biggs1975}, introducing an effective dispersion comparable to scattering on a 10-eV free-electron plasma \citep{Sunayev1996,XRISMCircinus2026}.} we obtain an upper limit of $11\%$ of the core flux. 

\begin{figure}[ht]
    \centering
        {\includegraphics[width=\columnwidth]{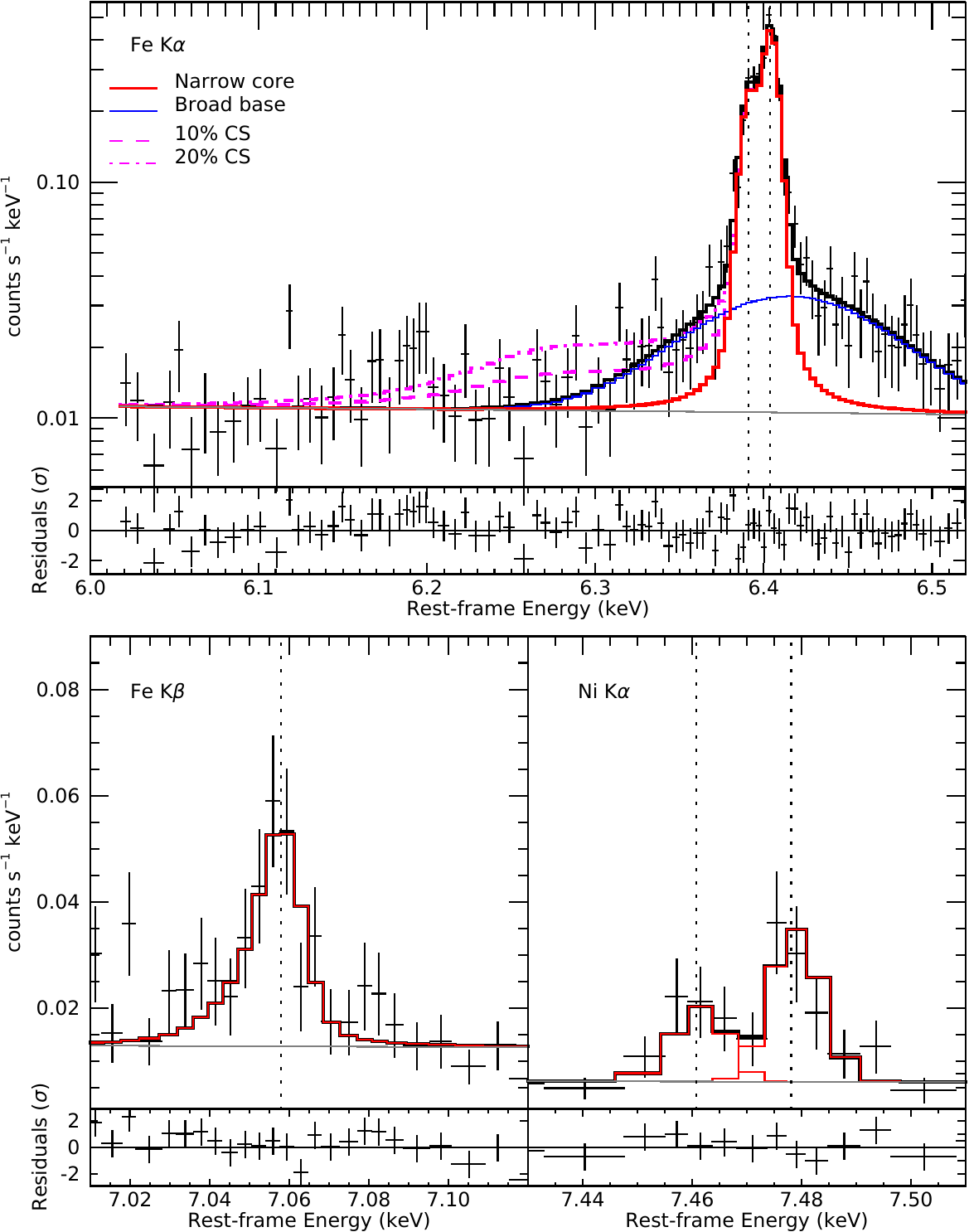}}
     \caption{Local fits to the neutral fluorescence lines in the XRISM/Resolve spectrum of NGC~1068. Top: Fe~K$\alpha$ complex (6.0--6.5~keV, rest frame). The black points show the data and the black histogram is the best-fitting model, consisting of a narrow core (red) and a broad base (blue; see text) added to the baseline continuum (gray). The magenta dashed curves show bound electron-scattering Compton-shoulder profiles with fluxes of 10 and 20\% of the core. Bottom left: Fe~K$\beta$ band. Bottom right: Ni~K$\alpha$ band. The vertical dotted lines mark the nominal centroid energies of the neutral transitions.
}
      \label{NGC1068_Neutral}
\end{figure}

A very significant improvement of $\Delta C=63$ for 2 degrees of freedom (d.o.f.) is instead obtained by adding another \texttt{zbfeklor}, which reproduces a much broader ($\sigma=2400\pm400$ km s$^{-1}$) base of the line. A further modest improvement ($\Delta C=8.5$ for 1 d.o.f.) comes from allowing the broad base to be shifted by $-700^{+250}_{-300}$ km s$^{-1}$, confirming that it mostly corrects the blue wing of the line, with less effect on the red wing. The flux of the narrow core is $5.8\pm0.2\times10^{-5}$ photons cm$^{-2}$ s$^{-1}$, corresponding to an equivalent width (EW) of $710\pm20$ eV with respect to the local continuum; the flux of the broad component is $0.39\pm0.06$ that of the narrow core. The overall Fe K$\alpha$ line is well reproduced by this model, with no obvious residuals, except for a weak resolved feature at $\sim6.19$ keV, barely above the 99\% detection limit (see upper panel in Fig.~\ref{NGC1068_Neutral}).

As a further step, we performed a combined fit with the neutral Fe K$\beta$ and Ni K$\alpha$ lines, adding the energy ranges $\sim7-7.1$ and $\sim7.4-7.5$ keV. In all three local fits the baseline continuum model is a power law with $\Gamma=1$ and a free independent normalization. The Fe K$\beta$ line is modeled with a \texttt{zbfekblor} component, which is a four-Lorentzian approximation to the line profile with energies, widths, and relative amplitudes as taken from \citet{Holzer1997}. The Ni K$\alpha$ line is modeled with two \texttt{zvgaussian} components, corresponding to the K$\alpha$ doublet, with energies fixed at 7.4781 and 7.4608 keV and a fixed intensity ratio of 2:1, since an experimental line profile for Ni is not available in \texttt{XSPEC}. The velocity shifts for Fe K$\beta$ and Ni K$\alpha$ are $30\pm 50$ and $20^{+60}_{-50}$ km s$^{-1}$, respectively. The velocity broadening is essentially unconstrained for Fe K$\beta$, while for Ni K$\alpha$ we obtain $\sigma = 150^{+70}_{-60}$ km s$^{-1}$. Given these results, we tied both the velocity shift and broadening to be the same for the Fe K$\alpha$ core, Fe K$\beta$, and Ni K$\alpha$ lines, resulting in a good fit ($C=138/129$ d.o.f.; see Fig.~\ref{NGC1068_Neutral}), with a common $\sigma=155^{+18}_{-16}$ and $v=-30\pm12$ km s$^{-1}$. The Fe K$\beta$/K$\alpha$ (core) ratio is $0.103\pm0.016$, while the Ni K$\alpha$/Fe K$\alpha$ (core) ratio is $0.066\pm0.012$. When a broad component with the same width as that measured for the broad Fe~K$\alpha$ line is also included  for Fe~K$\beta$ and Ni~K$\alpha$, the resulting upper limits on the broad-to-core flux ratio are significantly higher than the value observed for Fe~K$\alpha$, and therefore do not provide a meaningful constraint. We note that the Fe K$\beta$ line is contaminated by the blue wing of the broad \ion{Fe}{xxvi} feature (see next section). When this contamination is modeled, the inferred Fe K$\beta$ flux decreases by approximately 15\%, remaining consistent within 1$\sigma$ with the ratio reported here.

\subsection{\label{sec:ionized}The ionized medium}

\begin{figure}[ht]
    \centering
        {\includegraphics[width=\columnwidth]{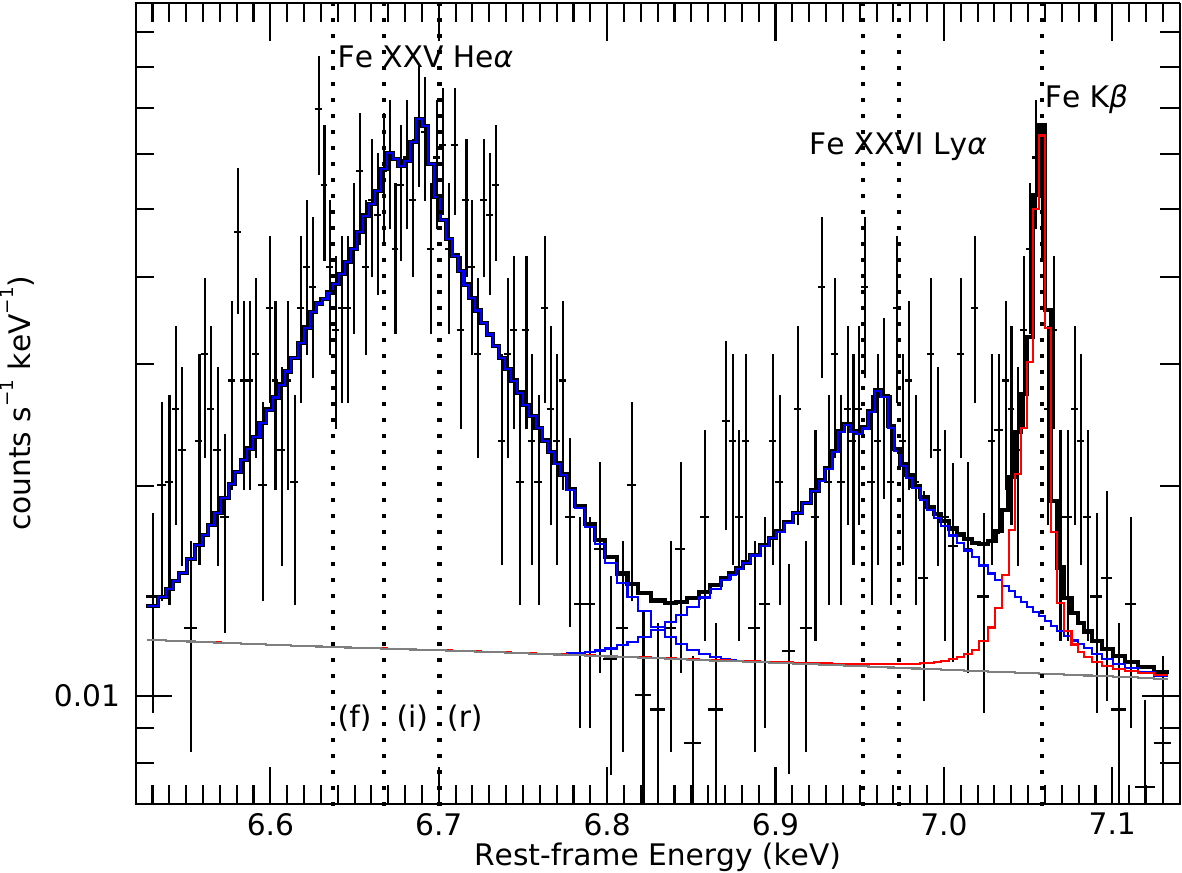}}
     \caption{XRISM/Resolve spectrum of NGC~1068 with the broad \ion{Fe}{xxv} He$\alpha$ and \ion{Fe}{xxvi} Ly$\alpha$ emission lines. The spectrum is shown in the rest frame and is fitted with the outflow model described in Sect.~4.2.1. The model components for the \ion{Fe}{xxv} and \ion{Fe}{xxvi} lines are shown in blue; the total model is shown in black. The neutral Fe~K$\beta$ line, shown in red, is also included in the fit. The vertical dotted lines mark the laboratory energies of the main transitions.
}
      \label{NGC1068_FeXXV}
\end{figure}

We performed a local fit in the $\sim6.5-7.1$~keV band, adopting a $\Gamma=1$ power law as the baseline continuum and including the Fe K$\beta$ emission line modeled in the previous section. Two broad Gaussian lines are required by the data, with rest-frame centroid energies of $6.674\pm0.004$ and $6.980\pm0.010$~keV and widths of $\sigma=2650^{+300}_{-200}$ and $2700\pm500$~km~s$^{-1}$. These features are readily identified as emission from \ion{Fe}{xxv} and \ion{Fe}{xxvi} (see Fig.~\ref{NGC1068_FeXXV}). Narrow residuals remain at $\sim6.56$ and $\sim6.605$~keV (observed), but modeling them as emission lines is not significant at the 99\% c.l. Including narrow Gaussian components at these energies does not significantly alter the properties of the broad lines in this phenomenological fit, nor does it affect the parameters derived from the models presented in the following sections, which remain consistent within the uncertainties.

We therefore modeled the \ion{Fe}{xxvi} feature as the Ly$\alpha$ doublet, with rest-frame energies fixed at 6.973 and 6.952~keV, a fixed intensity ratio of 2:1,\footnote{This ratio is strictly valid only in the optically thin limit; the column density inferred in Sect.~\ref{physical} is consistent with this regime, supporting the validity of this assumption \citep[e.g.,][]{Gunasekera2025}.} and a common velocity width and shift. The \ion{Fe}{xxv} He$\alpha$ complex was modeled with four narrow components (\emph{w, x, y, z}) at 6.700, 6.682, 6.668, and 6.637~keV,\footnote{Satellite lines are likely to contribute significantly \citep[see, e.g.,][]{Bianchi2005,Mochizuki2025}, but it is impossible to disentangle them in a blended broad feature in the current data.} tying their velocity width and shift to those of the \ion{Fe}{xxvi} doublet. The relative intensities within the \ion{Fe}{xxv} multiplet are basically unconstrained; for reference we report the standard diagnostic ratios $R\equiv z/(x+y)=0.5^{+0.8}_{-0.3}$ and $G\equiv(x+y+z)/w>0.2$. Any centroid shift is also poorly constrained, with $v=-300^{+300}_{-500}$~km~s$^{-1}$, while the common width is $\sigma=2400\pm200$~km~s$^{-1}$. The fit is statistically equivalent to the previous two-Gaussian fit, with $C=132/132$ d.o.f.

Local fits were also performed in the range $\sim2-2.7$ keV to model the \ion{Si}{xiv}, \ion{S}{xv}, and \ion{S}{xvi} lines. For the Ly$\alpha$ transitions of \ion{S}{xvi} and \ion{Si}{xiv}, we adopted doublets with fixed 2:1 intensity ratios, using NIST \citep{NIST_ASD} atomic energies of $E_1 = 2.6227$~keV and $E_2 = 2.6197$~keV, and $E_1 = 2.00608$~keV and $E_2 = 2.00433$~keV, respectively. The best-fit widths and shifts are $\sigma = 730^{+260}_{-180}$~km~s$^{-1}$ and $v = 80^{+180}_{-190}$~km~s$^{-1}$ for \ion{S}{xvi}, and $\sigma = 950\pm250$~km~s$^{-1}$ and $v = -450\pm200$~km~s$^{-1}$ for \ion{Si}{xiv}. For \ion{S}{xv} He$\alpha$ we used a triplet at 2.4304, 2.448, and 2.4606 keV with free relative normalizations, obtaining $\sigma = 690^{+240}_{-200}$~km~s$^{-1}$ and $v = -60^{+200}_{-190}$~km~s$^{-1}$. When all the lines are fitted simultaneously, we obtain a common width of $\sigma = 710^{+150}_{-120}$~km~s$^{-1}$ and a shift of $v = -100\pm140$~km~s$^{-1}$, with $C=32/25$ d.o.f.

\section{Discussion}

\subsection{The cold reflector}
\subsubsection{Location of the line-emitting region}
Assuming that the observed Gaussian broadening of the neutral fluorescent lines traces predominantly Keplerian motion at a characteristic emitting radius $r_{\rm em}$, we parameterize  the line-of-sight velocity dispersion as $\sigma_{\rm obs}=f_{\rm geo}\,v_{\rm K}\sin i$, where $v_{\rm K}=(GM_{\rm BH}/r_{\rm em})^{1/2}$ is the Keplerian speed, $i$ is the inclination of the rotating structure with respect to the line of sight, and $f_{\rm geo}$ accounts for geometric effects. Adopting $M_{\rm BH}=6.7\times10^{6}\,M_\odot$ \citep{Padovani2024} and $\sigma_{\rm obs}=155$ km s$^{-1}$ (see Sect.~\ref{neutrallines} and Table~\ref{tab:lines}), this yields $r_{\rm em}\simeq1.2\,(f_{\rm geo}\sin i)^2$ pc. We can compare this radius to the dust sublimation radius, estimated using the generic scaling $r_{\rm sub}=A_{\rm sub}\sqrt{L_{45}}~{\rm pc}$, where $L_{45}\equiv L_{\rm bol}/10^{45}\,{\rm erg\,s^{-1}}$ and $A_{\rm sub}$ is a normalization that depends on grain properties and sublimation temperature \citep{Barvainis1987, Nenkova2008}. Combining the two, we obtain
\begin{equation}
\label{eq:rsub}
\begin{aligned}
\frac{r_{\rm em}}{r_{\rm sub}} \sim\;
& 4.2\,
\left(\frac{A_{\rm sub}}{0.4}\right)^{-1}
\left(\frac{M_{\rm BH}}{6.7\times10^{6}\,M_\odot}\right)
\left(\frac{\sigma_{\rm obs}}{155~{\rm km\,s^{-1}}}\right)^{-2} \\
& \times
\left(\frac{L_{\rm bol}}{10^{44.7}~{\rm erg\,s^{-1}}}\right)^{-1/2}
(f_{\rm geo}\sin i)^2 ,
\end{aligned}
\end{equation}

\noindent where the numerical factor is evaluated for our fiducial $M_{\rm BH}$ and
$L_{\rm bol}$ \citep{Padovani2024}, the measured $\sigma_{\rm obs}$ reported above, and $A_{\rm sub}=0.4$, as valid for a sublimation temperature of 1500 K \citep{Barvainis1987, Nenkova2008}.

Independent infrared and submillimetre observations of NGC~1068 directly resolve dusty structures over a similar range of radii: near-IR interferometry with VLTI/GRAVITY locates the hottest dust at $r\simeq 0.2$~pc, commonly identified with the dust sublimation region \citep{Gravity2020,Vermot2021}, while mid-IR interferometry with VLTI/MIDI requires warm dust components extending over a few parsecs \citep{Raban2009,LopezGonzaga2014}, and ALMA observations resolve a
compact molecular and dusty structure with a characteristic
diameter of $\sim 7$~pc \citep{GarciBurillo2016,GarciaBurillo2019}. Thus,
the radius inferred from the Fe K$\alpha$ line width lies within the same
order of magnitude as the directly observed dusty structures in
NGC~1068, and is only a factor of a few larger than the hot-dust
radius associated with the sublimation zone.

Incidentally, from Eq.~\ref{eq:rsub} $r_{\rm em}=r_{\rm sub}$ requires
$f_{\rm geo}\sin i\simeq 0.49$. Two limiting cases are illustrative.
If $f_{\rm geo}\simeq1$ (i.e., the measured Gaussian width closely tracks
the full projected Keplerian speed scale), then $\sin i\simeq0.49$
($i\simeq29^\circ$), which is inconsistent with the nearly edge-on orientation inferred for NGC~1068 from independent constraints based on NLR kinematics, H$_2$O maser disk modeling, and mid-infrared interferometry  \citep[e.g.,][]{das_kinematics_2006,Greenhill1996,Raban2009}. Conversely, if a dust-sublimation bounded structure is close to edge-on ($\sin i\simeq1$), then $f_{\rm geo}\simeq0.5$, as might be expected for a rotation-dominated, flattened, or ring-like structure.

The emission radius inferred from the width of the Fe~K$\alpha$ core indicates that most of the line is produced in the inner circumnuclear region, on scales on the order of 1 parsec, although it does not exclude that a fraction of the emission arises at larger distances, contributing to a narrower profile. Consistently, the Resolve Fe~K$\alpha$ core flux is in agreement with that measured in the \textit{Chandra} HETG spectra by \citet{Ogle2003} and \citet{Kallman2014}, whose extractions encompass the central few arcseconds, corresponding to physical scales of order 100--200~pc (1\arcsec\ $\simeq$ 49~pc). \textit{Chandra} observations nevertheless also provide evidence for Fe~K$\alpha$ emission on larger scales: \citet{Ogle2003} detected weaker off-nuclear fluorescence, while \citet{Kallman2014} showed that the measured line flux depends on the extraction width. A more explicit decomposition was presented by \citet{Bauer2015}, who separated a compact ($<2\arcsec$) and an extended (2\arcsec--75\arcsec) component, also using  ACIS-S data, and found that about 30\% of the Fe~K$\alpha$ core flux originates outside the central 2\arcsec \citep[but see][for evidence that this value may be overestimated and should be regarded as an upper limit when pile-up effects are properly accounted for]{Andonie2022}. The intrinsic complexity of the line profile, as further highlighted by the Resolve spectrum, particularly in the modeling of the broad base, limits a more precise determination of the spatially extended fraction.

\subsubsection{Ionization state and elemental abundances}

\begin{figure}[ht]
    \centering
               {\includegraphics[width=\columnwidth]{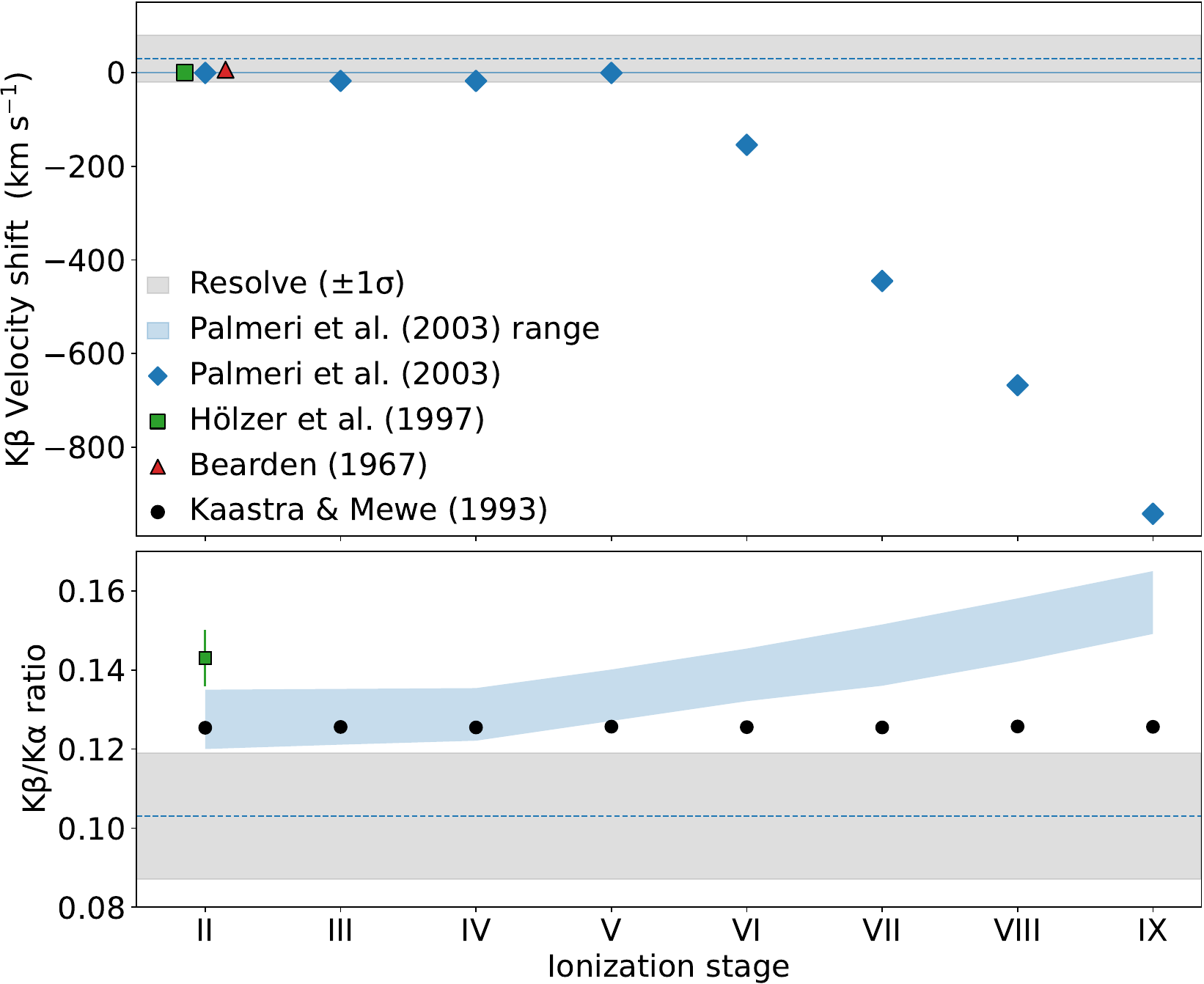}}
     \caption{\textit{Upper panel:} Velocity shifts of the Fe~K$\beta$ centroid energy relative to the neutral reference of \citet{Holzer1997}, shown as a function of ionization stage. Blue diamonds show the shifts predicted by the atomic calculations of \citet{Palmeri2003}, while the gray horizontal band indicates the value measured in the XRISM/Resolve spectrum ($\pm 1\,\sigma$). The green square and red triangle mark the experimental neutral Fe~K$\beta$ centroid energies from \citet{Holzer1997} and \citet{Bearden1967}, respectively. The measured centroid is consistent with neutral iron and rules out, within the statistical uncertainties, ionization stages higher than \ion{Fe}{v}. \textit{Lower panel:} Fe~K$\beta$/K$\alpha$ ratio as a function of ionization stage. The black circles show the values predicted by \citet{KaastraMewe1993}, while the light blue shaded region indicates the envelope spanned by the theoretical calculations of \citet{Palmeri2003}.
 The green square marks the experimental neutral Fe~K$\beta$/K$\alpha$ ratio from \citet{Holzer1997}, and the horizontal band indicates the ratio measured in the XRISM/Resolve spectrum ($\pm 1\,\sigma$). The observed ratio is consistent, within the uncertainties, with the theoretical expectation for neutral Fe in the optically thin limit, although marginally lower than the nominal value.
}
      \label{fig:FeKbKa}
\end{figure}

The upper panel of Fig.~\ref{fig:FeKbKa} shows the centroid energy of the Fe~K$\beta$ line as a function of ionization stage, compared with the value measured in the \emph{XRISM}/Resolve spectrum. Experimental measurements of the Fe~K$\beta$ centroid for neutral iron by \citet{Bearden1967} and \citet{Holzer1997} are in good agreement with the theoretical values computed by \citet{Palmeri2003}, indicating that the absolute reference energy of the line is robust. The observed centroid is consistent with iron in ionization states lower than \ion{Fe}{v}, while \ion{Fe}{vi} and higher ionization stages are excluded at the $\gtrsim 3\sigma$ level, as they would predict centroid shifts significantly larger than observed \citep{Palmeri2003,Nagai2026}. Although the Fe~K$\alpha$ line is measured with smaller statistical uncertainties, its diagnostic power is limited by significantly larger uncertainties in the expected centroid energies. Experimental measurements of neutral Fe~K$\alpha$ by \citet{Bearden1967} and \citet{Holzer1997} are mutually consistent, whereas theoretical calculations by \citet{Palmeri2003} predict different centroid energies for the individual K$\alpha_1$ and K$\alpha_2$ components (corresponding to offsets of $\sim-70$ and $\sim-30$~km~s$^{-1}$, respectively), reflecting the complex line profile produced by many closely spaced transitions. The same calculations also predict a redshift of the Fe~K$\alpha$ centroid with increasing ionization, at least in the first stages of ionization; for example, the shift from the neutral value to \ion{Fe}{vi} corresponds to a velocity offset of $\sim$150~km~s$^{-1}$. Instead, the XRISM spectrum shows a blueshift with respect to the neutral reference. This again argues against a significantly ionized origin for the fluorescent emission and suggests that any measured blueshift is more naturally interpreted as being due to gas kinematics rather than ionization effects.

The lower panel of Fig.~\ref{fig:FeKbKa} shows that the observed Fe~K$\beta$/K$\alpha$ ratio is lower but marginally consistent within $1\sigma$ with the range predicted for neutral iron in the optically thin limit by current atomic calculations\footnote{Accounting for contamination from the blue wing of \ion{Fe}{xxvi} reduces the Fe~K$\beta$ flux by $\sim$15\% (Sect.~\ref{neutrallines}), hence the Fe~K$\beta$/K$\alpha$ ratio, slightly increasing the tension with the neutral optically thin expectation.} (e.g., \citealt{KaastraMewe1993,Palmeri2003}; however, we note the higher ratio reported by \citealt{Holzer1997}). In the calculations of \citet{KaastraMewe1993}, the K$\beta$/K$\alpha$ ratio shows no significant variation with ionization stage at least up to \ion{Fe}{ix}. In contrast, the calculations of \citet{Palmeri2003} predict a marked increase in  the ratio at intermediate ionization stages, around \ion{Fe}{iv}--\ion{Fe}{v}. A decrease in the K$\beta$/K$\alpha$ ratio is expected only at substantially higher ionization stages, which are already robustly excluded by the centroid-energy constraints discussed above. We also investigated the effect of optical depth on the expected Fe~K$\beta$/K$\alpha$ ratio by computing the line emission in the optically thick, semi-infinite slab limit using the analytical formalism of \citet{Basko1978}, adopting photoelectric cross sections from \citet{Verner1996} and fluorescent yields from \citet{KaastraMewe1993}. The incident continuum was modeled as a power law with photon index $1.7 \leq \Gamma \leq 2.2$, over which the angle-averaged Fe~K$\beta$/K$\alpha$ ratio shows only a weak dependence on $\Gamma$. In this regime, reflection from Compton-thick material is expected to enhance the Fe~K$\beta$/K$\alpha$ ratio with respect to the optically thin case by a factor of $\sim8-15\%$ (depending on the inclination, and including only unscattered photons) because the photoelectric opacity decreases with energy, allowing K$\beta$ photons to escape more efficiently than K$\alpha$ photons. This increases the discrepancy with the observed value rather than alleviating it. Taken together, the line centroid energies and relative line intensities consistently point to reflection from predominantly neutral iron close to the optically thin limit, indicating that, at the parsec-scale distances inferred above, the emitting material is not compatible with a homogeneous Compton-thick reflector and instead arises mostly in lower column density gas.

The observed Ni~K$\alpha$/Fe~K$\alpha$ photon-flux ratio measured in the XRISM/Resolve spectrum of NGC~1068 is $0.066\pm0.012$. In the optically thin limit, \citet{Yaqoob2011} predict a Ni~K$\alpha$/Fe~K$\alpha$ ratio of 0.033 for $\Gamma=2$, with a very weak dependence on photon index and none on geometry or covering factor. This implies a nickel overabundance, with respect to \citet{Anders1989}, by a factor of $2.0\pm0.4$.
Using the same optically thick slab calculations described above, we also computed the Ni~K$\alpha$/Fe~K$\alpha$ core-to-core line ratio in the Compton-thick limit. As for iron, the angle-averaged Ni/Fe ratio shows only a very weak dependence on $\Gamma$, and exceeds the optically thin prediction by only about 2\%, leaving unchanged the required nickel overabundance. Evidence for a nickel overabundance in NGC~1068 was already suggested by previous studies based on lower-resolution data \citep{Matt2004}, with similar indications also reported for the Circinus galaxy \citep{Molendi2003,XRISMCircinus2026}.

\subsubsection{\label{sec:CS}The Compton shoulder and reflector column density}

We conclude this section with a particularly striking result that emerged from the analysis reported in Sect.~\ref{neutrallines}, namely the apparent weakness of the CS associated with the neutral Fe~K$\alpha$ line. Given the long-standing classification of NGC~1068 as a prototypical Compton-thick Seyfert~2, and the presence of a strong narrow Fe~K$\alpha$ core, a prominent CS would naturally be expected. Instead, the data reveal no clear evidence for a substantial red Compton wing, allowing us to place an upper limit of $\sim$8--11\% (depending on the adopted shape) on the CS flux relative to the line core.

In standard reflection scenarios in which the same homogeneous optically thick medium both obscures the nucleus and produces the bulk of the fluorescent emission, Monte Carlo calculations predict CS-to-core ratios of order $\gtrsim$15--20\%, depending on geometry and inclination \citep[e.g.,][]{Matt2002,Yaqoob11b}. The significantly weaker CS inferred here therefore disfavors a scenario in which the bulk of the observed Fe~K$\alpha$ emission arises in a uniform, classical Compton-thick reflector. This conclusion is fully consistent with the independent constraints derived above from the Fe~K$\beta$/Fe~K$\alpha$ ratio, which already disfavor very large effective column densities in the line-emitting region. It is further supported by the depth of the Fe~K absorption edge, for which we measure an optical depth $\tau_{\rm K} = 0.85 \pm 0.15$. This latter constraint is more sensitive to the adopted continuum modeling within the simplified phenomenological fits presented here, and should therefore be treated with caution; nevertheless, it remains broadly suggestive of a reflector of moderate effective column density.

A natural interpretation is that absorption and reflection are governed by different characteristic column densities. In this scenario, the line-of-sight obscuration in NGC~1068 is persistently Compton-thick, consistent with an inner torus that reaches column densities high enough to be self-obscuring, so that reflection produced at small radii is itself strongly attenuated along our line of sight. As a result, emission from the highest column-density regions remains largely hidden, and the observed time-averaged reflected spectrum is dominated by more extended reprocessing material, as also indicated by mid-infrared interferometric studies \citep[e.g.,][]{Raban2009} and by the evidence for a spatially extended component of the Fe K$\alpha$ emission discussed above. This naturally dilutes the observed CS relative to the line core. A quantitative assessment of this picture, however, requires a comprehensive, self-consistent treatment of the full reflection spectrum. This will be addressed in a future companion paper, based on joint modeling of the XRISM/Resolve spectrum together with the simultaneous Xtend and \emph{NuSTAR} data, using radiative transfer simulations \citep[e.g., SKIRT;][]{VanderMeulen2023} to robustly constrain the geometry and column-density distribution of the cold reflector.

\subsection{A warm bipolar outflow scenario}

High spatial resolution HST spectroscopy, together with MUSE integral-field spectroscopy and 3D kinematic modeling, has revealed that the NLR of NGC~1068 hosts very fast ionized gas components, with [\ion{O}{iii}]~$\lambda5007$ clouds reaching deprojected velocities of up to $\sim$2000~km~s$^{-1}$ while flowing within a well-defined biconical outflow structure \citep[e.g.,][]{crenshaw_resolved_2000,das_kinematics_2006,Marconcini2025}. When the emission from the entire NLR is integrated, the superposition of clouds spanning this wide velocity field naturally produces a broad [\ion{O}{iii}] line profile, as explicitly illustrated in Fig.~4 of \citet{crenshaw_radial_2010}. The same phenomenology is observed at higher ionization: recent JWST integral-field spectroscopy has demonstrated that the [\ion{O}{iv}]~26~$\mu$m emission traces the same global biconical outflow geometry inferred from the optical data, but requires a systematically faster kinematic component, with a deprojected maximum velocity larger by $\sim$300~km~s$^{-1}$ with respect to [\ion{O}{iii}] \citep{Marconcini2025}. These results indicate that higher-ionization lines preferentially sample dynamically hotter phases of the same outflow. In this context, the broad \ion{Fe}{xxv} and \ion{Fe}{xxvi} emission lines observed with XRISM/Resolve can be naturally interpreted as arising from an even more highly ionized component of the same large-scale outflow, whose kinematic signatures are integrated over the full extent of the emitting region.

\subsubsection{Model setup and comparison with the data}
\label{biconicalmodel}
\begin{figure}[ht]
    \centering
               {\includegraphics[width=\columnwidth]{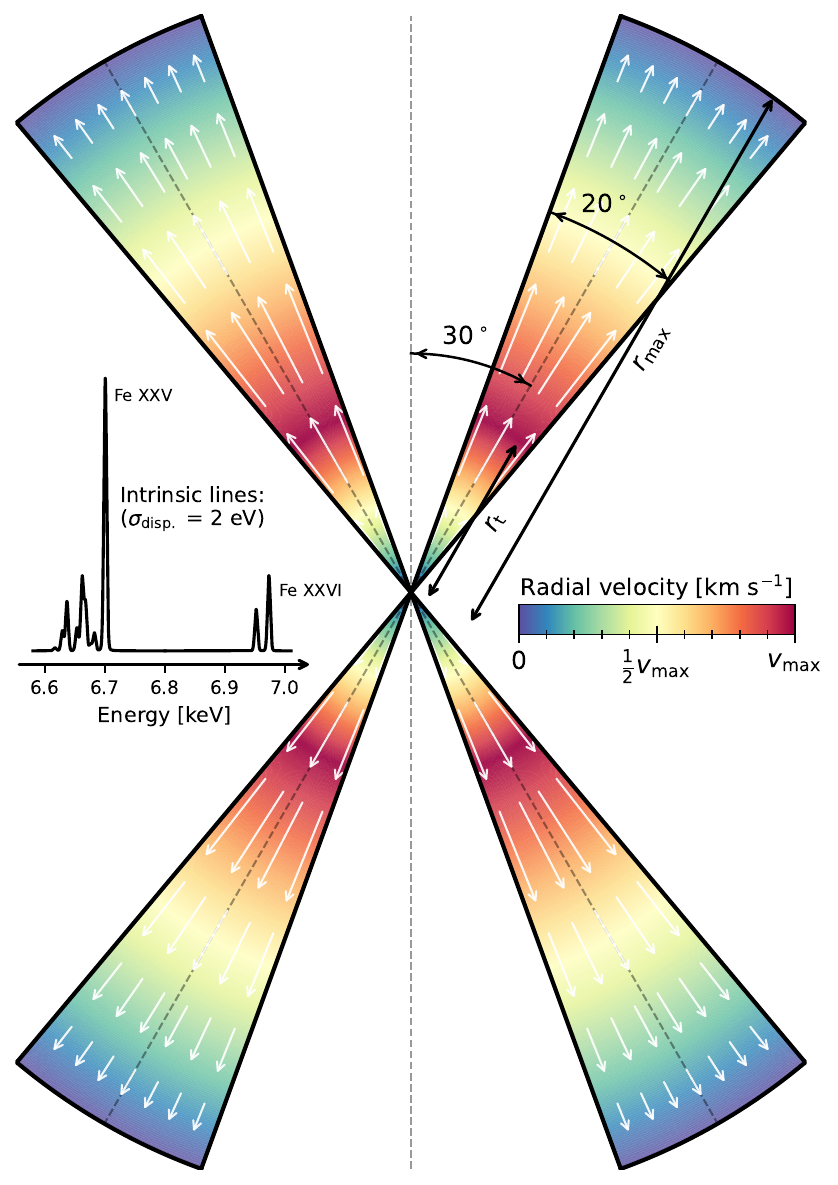}}
     \caption{High-velocity bi-polar outflow model adopted to reproduce the broad \ion{Fe}{xxv} and \ion{Fe}{xxvi} lines observed with XRISM. The intrinsic line spectrum was calculated with XSTAR; the integrated velocity profile was calculated with the SKIRT code.}
      \label{fig:biconicalmodel}
\end{figure}
\noindent
We assume that the broad \ion{Fe}{xxv} and \ion{Fe}{xxvi} lines are emitted directly by a high-velocity outflow in the polar direction, extending above and below the obscuring torus. Following previous modeling of the optical [\ion{O}{iii}] and infrared [\ion{O}{iv}] lines in NGC~1068 \citep{das_kinematics_2006, Marconcini2025}, we adopted a biconical outflow with a half-opening angle of $30^\circ$ and an angular width of $20^\circ$ (see Fig.~\ref{fig:biconicalmodel}), viewed at an inclination of $i=85^\circ$. The radial velocity field is called the Hubble flow, and has a velocity magnitude that varies with radial distance \citep{Das2005, das_kinematics_2006}. From $r = 0$ up to the turnover radius $r_\text{t}$, the velocity increases linearly from zero to $v_\text{max}$. Beyond $r_\text{t}$, the velocity decreases linearly from $v_\text{max}$ down to zero at $r = r_\text{max}$.

The intrinsic line spectra were computed with \textsc{xstar} \citep{Kallman2001} (similar intrinsic spectra are obtained with \textsc{cloudy}, last described by \citealt{Gunasekera2025a}), assuming photoionized gas illuminated by a generic AGN spectral energy distribution, and adjusted such that, after convolution with a phenomenological broadening model, the profiles reproduce the observed line shapes.
 We applied an additional Gaussian broadening of $2~\text{eV}$ ($\sim90$ km s$^{-1}$ at these energies) to the intrinsic line spectrum, consistent with the local velocity dispersion measured with JWST \citep{Marconcini2025}, which is much smaller than the broadening observed in the \ion{Fe}{xxv} and \ion{Fe}{xxvi} line wings with XRISM. The radial emissivity profile of the biconical outflow is proportional to $r^{-2}$, as expected for radiation-compressed, ionization-bounded gas \citep[e.g.,][]{Stern2014}. With this choice, the model is scale-invariant, and the predicted line profile is insensitive to the absolute values of $r_t$ and $r_{\max}$, depending only on the velocity field and geometry of the biconical outflow (see also Sect.~\ref{physical}).

The integrated velocity profile corresponding to this model was calculated with the SKIRT code \citep{Camps2015, camps2020}, adopting a \texttt{ConicalShellGeometry} source with a \texttt{HubbleRadialVectorField}\footnote{In the context of this work, we implemented the Hubble flow vector field in SKIRT; see \url{https://skirt.ugent.be/skirt9/class_hubble_radial_vector_field.html}} velocity field. All model parameters were fixed to the optical [\ion{O}{iii}] model \citep{das_kinematics_2006}, except for $v_\text{max}$, which was varied between $2000$ and $18000~\text{km s}^{-1}$. The simulation results were converted into an \texttt{XSPEC} tabulated additive model, which can be directly fitted to the XRISM data.

The best fit using this model is equivalent to the ones presented in Section~\ref{sec:ionized}, with $C=138/134$ d.o.f., and is shown in Fig.~\ref{NGC1068_FeXXV}. The free parameters are a \ion{Fe}{xxvi}/\ion{Fe}{xxv} ratio of $0.28\pm0.05$, a common shift of $460\pm90$ km s$^{-1}$, and a common $v_\text{max}=11\,000\pm1\,000$ km s$^{-1}$. No improvement is found if we allow $v_\text{max}$ to be different for the two lines, while a marginal improvement ($\Delta C=5$ for 1 d.o.f. less) if the two velocity shifts are allowed to vary independently, with the velocity shift for \ion{Fe}{xxvi} being consistent with 0. We note that the derived $v_\text{max}$ is significantly higher than that found for the optical and the IR outflows. 

To explore potential degeneracies that could reduce the inferred $v_\text{max}$, we ran a large grid of SKIRT simulations, varying all parameters relative to the optical [\ion{O}{iii}] reference values (i.e., the turnover radius, half-opening angle, angular width, maximum radius, inclination, and radial emissivity profile). We find that none of these parameters are constrained by the data, and none can significantly reduce $v_\text{max}$. This remains true even when allowing for large intrinsic line broadening via Gaussian smoothing, while increasing the viewing angle from $85^\circ$ to $90^\circ$ leads only to a marginal reduction of the best-fit $v_\text{max}$. This behavior is expected since the velocity field is anchored to the spatially resolved, empirically derived HST model of \citet{das_kinematics_2006}, whereas the XRISM spectrum is sensitive only to the velocity distribution integrated over the polar outflow; different radial profiles with the same emissivity-weighted average velocity therefore produce essentially indistinguishable line shapes. Overall, this exercise should be regarded as a toy model meant to compare the X-ray emitting gas with the optical outflow: while the higher velocity is robust, the remaining model parameters are not significantly constrained by the present data.

\subsubsection{Consistency and physical implications}
\label{physical}

Having shown that the highly ionized Fe emission is broadly consistent with our outflow interpretation, we now perform a simple internal consistency check. For simplicity, we focus on the brightest \ion{Fe}{xxv} He-$\alpha$ line and assess whether a  biconical structure illuminated by the intrinsic AGN continuum can plausibly reproduce its observed luminosity, given the source's hard X-ray output.

To this end, we adopt the optically thin formalism of \citet{Matt1996} and \citet{Yaqoob2001}, in which the line luminosity is proportional to the
number of continuum photons absorbed in the Fe K shell, multiplied by an
effective fluorescent yield. We assume an AGN continuum described by a power law with photon index $\Gamma=2$, normalized to an intrinsic $2$--$10$~keV luminosity $L_{2-10}=10^{43.4}\,\mathrm{erg\,s^{-1}}$ \citep{Padovani2024}; the medium is optically thin to Fe K-shell photoabsorption and is characterized by a uniform hydrogen column density along all directions from the central source intersecting the emitting region. Within this approximation, the geometry enters only through the covering factor $C_f$, defined as the fraction of solid angle subtended by the gas as seen from the central source.

The observed luminosity of the \ion{Fe}{xxv} He-$\alpha$ complex is
$L_{\rm Fe\,XXV}=10^{39.75}\,\mathrm{erg\,s^{-1}}$. For He-like iron, line emission arises from recombination rather than fluorescence; we therefore adopt an effective fluorescent yield $Y_{\rm eff}=0.76$ \citep{Matt1996}, defined as the probability that the recombination cascade produces a K-shell photon, a K-shell threshold energy
$E_K=8.83$~keV and a photoionization cross section at threshold
$\sigma_K(E_K)=1.96\times10^{-20}\,\mathrm{cm^2}$ \citep{Verner1996}, with an $E^{-3}$ dependence above. The adopted iron abundance is taken from \citet{Anders1989}, and the fraction of iron in the He-like charge state is fixed at its peak of $f_{\rm XXV}=0.5$ \citep[e.g.,][]{Mehdipour2016}.

Under these assumptions, the ratio of the \ion{Fe}{xxv} line luminosity
to the intrinsic $2$--$10$~keV luminosity depends only on the product
$C_f\langle N_{\rm H}\rangle$, where $\langle N_{\rm H}\rangle$ denotes the
characteristic hydrogen column density encountered by rays intersecting the emitting region. Adapting and solving the \cite{Yaqoob2001} and \citet{Matt1996} equations yields
\begin{equation}
C_f\,\langle N_{\rm H}\rangle
=
5\times10^{21}\;\mathrm{cm^{-2}} .
\label{cnh_equation}
\end{equation}

For our hollow biconical geometry, defined by polar angles
$\theta=20^\circ$--$40^\circ$ from the symmetry axis, the covering factor is
$C_f=\cos20^\circ-\cos40^\circ=0.1736$. This implies an average hydrogen column density $\langle N_{\rm H}\rangle=2.9\times10^{22}
\;\mathrm{cm^{-2}}$, representative of the \ion{Fe}{xxv}-emitting gas along directions intersecting the bicone walls. We therefore conclude that, at the order-of-magnitude level probed by this photon-budget test, a bipolar cone illuminated by the intrinsic AGN continuum is fully capable of producing the observed \ion{Fe}{xxv} He-$\alpha$ luminosity. From this energetic standpoint, the proposed geometry is thus a viable and self-consistent scenario.

We emphasize that this photon-budget test accounts only for the
recombination-driven component of the \ion{Fe}{xxv} He-$\alpha$ complex.
As discussed by \citet{Matt1996} and \citet{Bianchi2005}, resonant
scattering can further enhance the \ion{Fe}{xxv} emission in photoionized
gas. Any such contribution would reduce the number of recombinations required and
thus lower the column density needed to reproduce the observed line
luminosity; the values of $C_f\langle N_{\rm H}\rangle$ derived above should
therefore be regarded as conservative upper limits on the required column. In any case, the observed large velocity spread is expected to reduce the efficiency of resonant scattering through Doppler de-saturation, while simultaneously suppressing resonant trapping and self-absorption that could otherwise reshape the line profile. The observed ionized Fe profiles are instead smooth and well reproduced by a simple biconical outflow model, indicating that these effects are not dominant.

We can extend this reasoning by combining the above column-density constraint
with the ionization requirements for \ion{Fe}{xxv}. The ionization parameter is defined as $\xi \equiv L_{\rm ion}/(n_{\rm H} r^2)$, where $L_{\rm ion}$ is the ionizing luminosity. For NGC~1068 we adopt a bolometric luminosity
$\log L_{\rm bol}=44.7$ \citep{Padovani2024}, and assume
$L_{\rm ion}\simeq\tfrac{1}{2}L_{\rm bol}$ following \citet{Panda2022}, yielding
$L_{\rm ion}\simeq2.5\times10^{44}\,\mathrm{erg\,s^{-1}}$. The \ion{Fe}{xxv} ionic fraction peaks at $\xi\simeq10^3~\mathrm{erg\,cm\,s^{-1}}$ \citep[e.g.,][]{Mehdipour2016}. Requiring this ionization state therefore implies that, if \ion{Fe}{xxv} is produced over an extended range of radii, the gas density must scale as
$n_{\rm H}\propto r^{-2}$. Combining this with the inferred column density, the characteristic radial thickness of the \ion{Fe}{xxv}-emitting layer at radius $r$ is
\[
\Delta R(r)\sim \frac{\langle N_{\rm H}\rangle}{n_{\rm H}(r)}
= \langle N_{\rm H}\rangle\,\frac{\xi\,r^2}{L_{\rm ion}},
\]
which increases rapidly with radius as $\Delta R\propto r^2$. Imposing the minimum geometrical requirement $\Delta R\lesssim r$ yields an upper bound on the radius at which a recombination-dominated \ion{Fe}{xxv} component can be sustained,
\begin{equation}
r \lesssim \frac{L_{\rm ion}}{\langle N_{\rm H}\rangle\,\xi}
\simeq 3\, \mathrm{pc}.
\label{r_equation}
\end{equation}

This indicates that, within a pure recombination scenario and assuming a roughly constant ionization parameter, the bulk of the \ion{Fe}{xxv} emission is naturally expected to originate on parsec scales. This scale is comparable to our estimate for the bulk of the neutral Fe~K$\alpha$ emission derived from its velocity width (see Eq.~\ref{eq:rsub}). In a classical unification picture, this would suggest that the ionized outflow is observed at radii comparable to those of the dominant neutral reflector, implying limited radial separation between the obscuring structure and the emitting gas \citep[e.g.,][]{Antonucci1993}. However, the neutral Fe~K$\alpha$ emission may also be partially or predominantly produced in a more extended medium, as suggested by the weak CS and by evidence for spatially extended reflection (see Sect~\ref{sec:CS}). In that case, the material responsible for blocking our direct view of the nucleus need not be co-spatial with the bulk of the neutral Fe~K$\alpha$ emitter, but may instead reside at smaller radii. In this geometry, the inner obscurer can remain sufficiently stable along the line of sight to avoid producing detectable variability below 10~keV, while changes in column density or covering factor affect only the hard X-ray transmitted component above 10~keV \citep{Marinucci2016,Zaino2020}. The ionized outflow can then be directly
observed without requiring a finely tuned patchy configuration.

Similarly to the neutral Fe~K$\alpha$ line, a comparison between the ionized Fe line fluxes measured by XRISM/Resolve and those reported from \textit{Chandra}/HETG provides constraints on the spatial origin of the ionized emission. For \ion{Fe}{xxvi}, the Resolve flux is consistent, within the uncertainties, with the HETG nuclear measurements reported by \citet{Ogle2003} and \citet{Kallman2014}, indicating that most of the emission is produced within the $\sim$100--200~pc nuclear region. In contrast, the \ion{Fe}{xxv} flux measured by Resolve is higher than that reported in the HETG nuclear spectra. This difference is likely related to the fact that the HETG analyses report unresolved line profiles, with upper limits on the FWHM that are significantly smaller than the widths measured by Resolve, so that part of the flux associated with the broadest velocity components may not be fully recovered. At the same time, there are indications that a fraction of the \ion{Fe}{xxv} emission may arise on larger spatial scales, as suggested by the HETG data presented by \citet{Kallman2014} and more explicitly quantified by \citet{Bauer2015}, who decomposed the emission into compact and extended components and found that a non-negligible fraction of the \ion{Fe}{xxv} flux may originate outside the central $\sim$100~pc. While these results do not exclude a contribution from extended ionized Fe emission, possibly associated with collisionally ionized gas on larger scales, they indicate that the bulk of the ionized Fe flux arises from a more compact region and can be naturally associated with the inner outflow, in agreement with the few-parsec limits derived above.

When placed in the context of the large-scale outflow, however, these parsec-scale
constraints are much smaller than those inferred for the [\ion{O}{iii}] biconical outflow, modeled with the same geometry adopted here, which yield
$r_t = 140$~pc and $r_{\max} = 461.9$~pc, as obtained by fitting a biconical kinematic model to the spatially resolved radial velocity field measured with \textit{HST} \citep{das_kinematics_2006}. From a kinematic point of view, the optical and infrared outflows already show a stratification with ionization state. In the biconical model of \citet{das_kinematics_2006}, the [\ion{O}{iii}] outflow reaches maximum deprojected velocities of $v_{\max}\simeq 2000$~km~s$^{-1}$. Using the same kinematic prescription, \citet{Marconcini2025} find that the mid-infrared
[\ion{O}{iv}]~25.9~$\mu$m emission reaches systematically higher velocities,
i.e., $\sim$300~km~s$^{-1}$ faster than [\ion{O}{iii}]. The ionized Fe emission
detected by XRISM/Resolve extends this trend to much more extreme velocities, with an inferred maximum outflow velocity of $v_{\max}\simeq 11\,000$~km~s$^{-1}$. These velocities exceed not only those measured in the optical and infrared, but also those of lower-ionization X-ray lines, such as S and Si, whose FWHM are approximately a factor of three narrower than those of \ion{Fe}{xxv} and \ion{Fe}{xxvi} (see Sect.~\ref{sec:ionized}). Other soft X-ray emission lines observed with \textit{Chandra}/HETG show velocity widths comparable to those of the Si and S lines and are spatially extended along the ionization cones \citep{Kallman2014}.
 Together, these measurements show that a kinematic stratification is already present within the X-ray band, with progressively higher velocities associated with higher-ionization states. The continuity of this trend from the optical and infrared to the hard X-rays is consistent with a spatially and kinematically stratified outflow across all wavebands.

An additional piece of evidence pointing toward a more complex circumnuclear
environment is provided by the neutral Fe~K$\alpha$ line, whose broad component exhibits a velocity width comparable to that of the ionized Fe emission. If confirmed, this would suggest that at least a fraction of the Fe~K$\alpha$-emitting material participates in the same dynamical environment as the highly ionized gas, for instance in the form of cold clumps embedded in the flow or fluorescence occurring in a compact, kinematically active region close to the nucleus. In any case, the interpretation of the Fe~K$\alpha$ profile requires further dedicated analysis, as part of the observed broadening may be due to the blending of emission from mildly ionized iron species \citep[e.g.,][]{Kallman2004,Bianchi2005b}.

\subsubsection{Energetics and feedback implications}

The discovery with XRISM/Resolve of a fast ionized outflow component
allows us to derive order-of-magnitude estimates of its mass, momentum, and
kinetic power by applying the standard scaling relations commonly used for AGN
winds, which relate the outflow energetics to its characteristic velocity,
column density, and radius \citep[e.g.,][]{Krongold2007}. Adopting the values inferred in \eqref{cnh_equation} and \eqref{r_equation}, $\mu_\mathrm{H}\simeq 1.4$ as the mean atomic weight per hydrogen atom \citep[as appropriate for solar abundances in][]{Anders1989}, and $v=5\,500\ \mathrm{km\,s^{-1}}$ (the observed FWHM is $\sim5600$ km s$^{-1}$ and the luminosity-weighted average velocity of our bipolar outflow model is v$_\mathrm{max}/2\simeq5500$ km s$^{-1}$; see Sect. \ref{sec:ionized} and \ref{biconicalmodel}), we obtain 

\begin{align}
\dot M &= 12
\left(\frac{C_f\,\langle N_{\rm H}\rangle}{5\times10^{21}\ {\rm cm^{-2}}}\right)
\left(\frac{r}{3\ {\rm pc}}\right)
\left(\frac{v}{5500\ {\rm km\ s^{-1}}}\right)
\, M_\odot\,{\rm yr^{-1}} ,\\
\dot E_{\rm k} &= 1.1\times10^{44}
\left(\frac{C_f\,\langle N_{\rm H}\rangle}{5\times10^{21}\ {\rm cm^{-2}}}\right)
\left(\frac{r}{3\ {\rm pc}}\right)
\left(\frac{v}{5500\ {\rm km\ s^{-1}}}\right)^3
\, {\rm erg\ s^{-1}}.
\end{align}

These estimates are subject to substantial systematic uncertainties, as they rely on simplifying assumptions on the geometry, ionization structure, and emissivity of the gas (Sect.~\ref{physical}), and in particular on the use of a single characteristic radius, column density, and velocity to describe a likely stratified outflow. The resulting kinetic power corresponds to a fraction of the bolometric luminosity of order $\dot{E}_{\rm k}/L_{\rm bol}\simeq 20\%$. Even adopting $L_{\rm bol}=L_{\rm Edd}$ as a conservative upper bound, the coupling would remain at the $\sim10\%$ level. These values are well above the typical thresholds of a few percent often invoked for AGN-driven feedback to be energetically significant \citep[e.g.,][]{DiMatteo2005,Hopkins2010}, indicating that this fast ionized component could play an important role in the coupling between the nuclear outflow and the surrounding medium. Although the spatial extent of the Fe-emitting gas is not directly resolved, it must originate outside the innermost obscured regions and is therefore likely distributed on subparsec scales to scales of a few parsecs. For characteristic radii of this order and velocities of several thousand km s$^{-1}$, the implied dynamical timescales are of order $t_{\rm dyn} \sim R/v \sim 10$--$10^3$ yr. Even allowing for substantial uncertainties in both geometry and energetics, such timescales suggest that the outflow need not be a purely transient phenomenon, and could lead to significant cumulative energy deposition in the circumnuclear environment. In this context, it is worth noting that NGC~1068 is believed to be accreting at a substantial fraction of its Eddington rate
($L_{\rm bol}/L_{\rm Edd}\sim 0.5$; \citealt{Padovani2024}), a regime in which powerful winds are expected (e.g., \citealt{King2015}), so the presence of a particularly powerful and efficient outflow in this source may not be unexpected.

To date, XRISM has revealed resolved broad ionized Fe emission in two other AGN, NGC~7213 and M81$^\ast$. In NGC~7213, the \ion{Fe}{xxv} and \ion{Fe}{xxvi} emission lines are detected with ${\rm FWHM}\simeq 3\,000$--$5\,000\ {\rm km\ s^{-1}}$, plausibly associated with an inner wind or dynamically hot plasma close to the nucleus (\citealt{Kammoun2025}; Murakami et al.\ in prep.). Similarly, M81$^\ast$ shows broadened \ion{Fe}{xxv}--\ion{Fe}{xxvi} emission with velocity widths of ${\sim}2\,000$--$3\,000\ {\rm km\ s^{-1}}$, together with centroid redshifts indicative of complex inner kinematics \citep{Miller2025}. Fast Fe--K emission has also been reported with \textit{Chandra} in the Compton-thick Seyfert~2 galaxy Mrk~34, where \citet{Maksym2023} detect extended \ion{Fe}{xxv}--\ion{Fe}{xxvi} emission associated with high-velocity ($\gtrsim 1.5\times10^{4}\ \rm km\ s^{-1}$) gas on large spatial scales. With the addition of NGC~1068, a coherent picture emerges in which broad ionized iron lines appear to be a common feature in nearby AGN spanning orders of magnitude in accretion rate, when observed at sufficient spectral resolution, tracing a previously underappreciated highly ionized wind component that may contribute significantly to AGN feedback.

\subsubsection{Alternative scenarios}
\label{sec:alternative}
The observed broadening of the ionized Fe lines does not necessarily require bulk outflow motions, as comparable velocity widths can arise from Keplerian motions in the broad-line region (BLR), as observed for optical permitted lines in type~1 AGN. In obscured sources, however, the BLR is not directly visible, but is detected only in polarized light, indicating that the direct view of the inner BLR is fully blocked and that the broad lines are observed after scattering away from the equatorial plane \citep{Antonucci1985}. In NGC~1068, these spectropolarimetric observations provided the foundational evidence for the AGN Unification Model \citep{Miller1991,Antonucci1994}, which explains the observed dichotomy between type~1 and type~2 Seyfert galaxies \citep{Antonucci1993,Urry1995,Netzer2015}. The hidden BLR in NGC~1068 exhibits broad permitted lines with observed FWHM of about $3000$--$3500~\mathrm{km~s^{-1}}$ in polarized light \citep{Antonucci1985,Miller1991}, comparable in order of magnitude to the velocity widths measured for the ionized Fe lines with XRISM.

We therefore adopt the same optical broad-line model, consisting of an optically thick equatorial torus and a static polar reflector (the UM mirror), observed at high inclination \citep{Miller1991, Ogle2003}. We assume that broad \ion{Fe}{xxv} and \ion{Fe}{xxvi} lines with a FWHM of $\sim5600$ km s$^{-1}$ are emitted from the inner BLR, which is fully obscured by the torus. We then simulate X-ray radiative transfer through this 3D geometry using the SKIRT code \citep{VanderMeulen2023}, which self-consistently treats photo-absorption, fluorescence, and scattering.

\begin{figure}[ht]
    \centering
               {\includegraphics[width=\columnwidth]{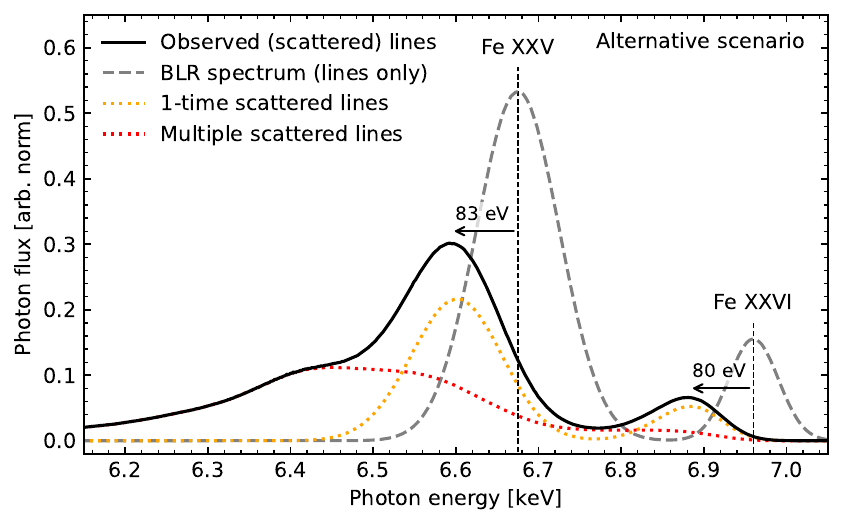}}
     \caption{Radiative transfer results for the scattering (UM mirror) scenario. The scattered \ion{Fe}{xxv} and \ion{Fe}{xxvi} lines are shifted by $\sim80$ eV due to inelastic Compton scattering, inconsistent with the observed line shift ($\Delta E < 15$ eV, consistent with zero).}
      \label{fig:UMmirror}
\end{figure}
The SKIRT radiative transfer results are shown in Fig.~\ref{fig:UMmirror}, demonstrating that the scattered \ion{Fe}{xxv} and \ion{Fe}{xxvi} emission lines would be shifted by $\sim80~\text{eV}$ relative to their intrinsic line energies. In the X-ray regime, electron scattering is inelastic because of Compton recoil: photons lose energy upon scattering by an amount that roughly scales as $E^2/m_{\rm e}c^2$ and depends on the scattering angle, making the effect substantial at Fe-K energies but negligible in the optical regime. The total energy shift depends on the number of scatterings (and therefore on the optical depth of the UM mirror), but even a single scattering produces shifts of $75~\text{eV}$ and $77~\text{eV}$ for \ion{Fe}{xxv} and \ion{Fe}{xxvi}, respectively. The XRISM Resolve spectrum rules out an energy shift $>15~\text{eV}$, conservatively derived from the outflow model (the phenomenological fit yields a shift consistent with zero), rejecting the hypothesis that the broad X-ray lines appear through scattering away from the equatorial plane.

A remaining possibility is that the broad ionized Fe lines arise from Keplerian motions in the inner BLR, and that we are observing this component directly through patchy obscuration. In this scenario, the equatorial absorber would remain optically thick at optical wavelengths, fully hiding the classical BLR, but could be partially transparent in hard X-rays, allowing a direct view of the rotating X-ray emitting region. To test this hypothesis, we convolved the same photoionized emission tables (\texttt{Cloudy}/\texttt{XSTAR}) adopted for the outflow with a blurring kernel (\texttt{rdblur} in \texttt{XSPEC}), representing emission from a rotating disk. This model provides a fit statistically comparable to that of the bipolar outflow scenario ($C = 138/136$), with an  inner radius $R_{\rm in}=150\pm70\,r_g$ and inclination $i = 15\pm1^\circ$, while the outer radius ($5000\,r_g$) and emissivity index ($q = -2$) were kept fixed.

Although this demonstrates that Keplerian rotation at BLR-like radii can phenomenologically reproduce the observed line width,  it implies a low inclination that is difficult to reconcile with the independently inferred, nearly edge-on geometry of NGC~1068  \citep[e.g.,][]{das_kinematics_2006,Greenhill1996,Raban2009}.
 More importantly, a direct view of such a compact, rotationally broadened component would be expected to produce detectable variability in the X-ray continuum and/or in the ionized Fe line emission on timescales much shorter than those already probed by past observations. The remarkable stability of the X-ray spectrum and Fe~K emission over more than two decades (e.g., \citealt{Bauer2015,Marinucci2016}) therefore disfavors a dominant contribution from a directly observed, rapidly varying BLR component. Future \textit{XRISM} observations will further test this possibility by searching for variability in the ionized Fe line fluxes and profiles.

A purely collisional ionization equilibrium scenario was also explored by fitting the ionized iron emission with an \texttt{APEC} model. This model provides a statistically comparable description of the data to the outflow interpretation, with $C=134$ for 135 d.o.f.. The best-fit temperature is $kT=6.1\pm0.4$~keV, and an emission measure ${\rm EM}=(5.4\pm0.6)\times10^{63}\,{\rm cm^{-3}}$. The fit still requires substantial line broadening, with a Gaussian velocity dispersion of $\sigma_v=2100\pm200$~km~s$^{-1}$. This is much larger than the thermal Doppler broadening expected for iron ions at this temperature, for which $\sigma_v^{\rm th}\simeq100$~km~s$^{-1}$.

Producing and sustaining such velocities in a hot, collisionally ionized plasma would most naturally require strong shocks or large-scale dynamical motions. In practice, shocks capable of accelerating gas to these velocities would imply a rapidly expanding or flowing medium, effectively reintroducing an outflow scenario \citep[e.g.,][]{Veilleux2005,King2015}. Similarly, invoking turbulence at this level would require extreme and sustained energy injection, with velocity dispersions comparable to the sound speed of a $kT\simeq6.1$~keV plasma. The emission measure implied by the \texttt{APEC} fit requires a substantial amount of hot gas, which for any reasonable characteristic size corresponds to a high thermal pressure. Combined with the bolometric thermal luminosity of the best-fit model, $L_{\rm th,bol}\simeq1.3\times10^{41}$~erg~s$^{-1}$, this implies that such a plasma would need to be continuously powered and would otherwise evolve on a short dynamical timescale ($t_{\rm dyn}\sim R/c_{\rm s}\approx8\times10^{2}\,(R/{\rm pc})$~yr for $kT\simeq6.1$~keV). While these considerations do not rule out this interpretation, and the AGN could in principle supply the required power, they indicate that a purely collisional origin does not naturally account for the combination of high ionization and large velocity widths, and therefore does not appear favored compared to interpretations invoking a dynamically active inner outflow.

We note, however, that strong shocks and highly turbulent gas are indeed observed in NGC~1068 on larger scales. Integral-field optical spectroscopy reveals very broad emission-line profiles in regions perpendicular to the ionization cones and radio jet, interpreted as shock-heated, turbulent gas produced by jet--ISM interaction \citep{Venturi2021}. In the same work, the \textit{Chandra} spectrum extracted from these extended, shock-dominated regions shows tentative evidence of ionized Fe emission. To further test this possibility, we fitted the XRISM spectrum with a non-equilibrium ionization shock model (\texttt{bpshock}). The fit converges toward parameters consistent with collisional ionization equilibrium, with an upper ionization timescale parameter $\tau_u \sim 5 \times 10^{13}\ \mathrm{s\,cm^{-3}}$, indicating that no significant non-equilibrium effects are required by the data. However, the XRISM spectrum integrates over a much larger region and is dominated by the bright nuclear component; any extended shock contribution would therefore be substantially diluted in the total spectrum.

\section{Conclusions}

In this paper we presented the first high-resolution X-ray microcalorimeter view of the iron-K emission in the prototypical Compton-thick Seyfert galaxy NGC~1068, obtained with XRISM/Resolve. Exploiting the unprecedented spectral resolution across the Fe~K band, we derived direct constraints on the ionization state, kinematics, and physical origin of the neutral and ionized iron emission. Our main results can be summarized as follows:

\begin{itemize}
\item The centroid energies of the Fe~K lines and the Fe~K$\beta$/K$\alpha$ flux ratio indicate that the fluorescent emission arises from genuinely neutral material and, together with the upper limit on the Compton shoulder ($\lesssim$8--11\% of the core flux), is dominated by optically thin or only mildly Compton-thick gas. This result is unexpected for a source long regarded as a prototypical Compton-thick AGN, although it is consistent with past evidence of significant clumpiness of the putative torus. The intrinsic width of the Fe~K$\alpha$ core ($\sigma \simeq 155$~km~s$^{-1}$) implies characteristic emission radii of order $\sim$1~pc, comparable to the directly resolved dusty structures, but likely distinct from the material responsible for the line-of-sight obscuration.

\item The \ion{Fe}{xxv} and \ion{Fe}{xxvi} emission lines are found to be remarkably broad, with Gaussian velocity dispersions of $\sigma = 2400 \pm 200$~km~s$^{-1}$. Their smooth, symmetric profiles are well reproduced by a biconical outflow model previously developed to explain the [\ion{O}{iii}] and [\ion{O}{iv}] emission, with the \ion{Fe}{xxv}--\ion{Fe}{xxvi} lines tracing a more highly ionized, much faster, and spatially more confined inner phase of the same large-scale bipolar outflow. The neutral Fe~K$\alpha$ profile also shows a broader base with a velocity width comparable to that of the ionized lines, possibly consistent with emission from cold clumps embedded in a multi-phase outflow.

\item Adopting fiducial parameters for the fast ionized phase (FWHM $\simeq 5600$~km~s$^{-1}$, $C_{\mathrm f}N_{\mathrm H} \simeq 5 \times 10^{21}$~cm$^{-2}$, and a characteristic radius of a few parsecs), the inferred mass outflow rate is of order $\sim$10~M$_\odot$~yr$^{-1}$, while the associated kinetic power reaches $\sim$10–20\% of the AGN bolometric luminosity, although these values are highly uncertain and may be overestimated due to the simplified treatment of the outflow structure and emissivity. This suggests that the inner, highly ionized outflow phase traced by the iron lines is energetically capable of providing significant AGN-driven feedback.
\end{itemize}

Overall, these results provide a revised view of the circumnuclear environment of NGC~1068, in which a clumpy and stratified cold reprocessor coexists with a fast, highly ionized inner outflow that carries a substantial fraction of the accretion power. Together with recent XRISM observations of other nearby AGN, this study shows that high-resolution X-ray spectroscopy is revealing a complex and structured picture of the innermost regions, in which multiple reprocessing and outflowing components with distinct ionization and kinematic properties coexist. Future XRISM observations will allow this emerging framework to be tested on a broader sample, while next-generation missions such as \textit{NewAthena} will eventually extend these studies to higher-redshift and lower-luminosity regimes.

\begin{acknowledgements}
We thank S.B. Kraemer, A. Laor and M. Tsujimoto for useful discussions.
ChatGPT (OpenAI) was used for language editing and code debugging during the preparation of this manuscript. All scientific content, analysis, and conclusions were independently developed and verified by the authors.
SB and CP acknowledge funding from PRIN MUR 2022 SEAWIND 2022Y2T94C, supported by European Union-Next Generation EU, Mission 4, Component 1, CUP C53D23001330006. BV acknowledges support through the European Space Agency (ESA) Research Fellowship Programme in Space Science. P.O.P. acknowledges financial support from the CNRS ``Action Thématique Phenomenes Extremes et Multimessangers'' (ATPEM) as well as from the French spatial agency CNES. VEG acknowledges funding under NASA contract 80NSSC24K1403. RS acknowledges funding from the CAS-ANID grant number CAS220016. GM acknowledges support from grant n. PID2023-147338NB-C21 funded by Spanish MICIU/AEI/10.13039/501100011033 and ERDF/EU. KI acknowledges support under the grant PID2022-136828NB-C44 provided by MCIN/AEI/10.13039/501100011033 / FEDER, UE. FN acknowledges support from the INAF-AF large grant X2X, ID 1.05.23.01.05 as well as the PRIN MUR 2022 DRAGON, ID 2022K9N5B4. GP acknowledges financial support from the European Research Council (ERC) under the European Union’s Horizon 2020 research and innovation program HotMilk (grant agreement No. 865637) and from the Framework per l’Attrazione e il Rafforzamento delle Eccellenze (FARE) per la ricerca in Italia (R20L5S39T9).
\end{acknowledgements}

\bibliographystyle{aa}
\bibliography{Bianchi26_NGC1068}

\begin{appendix}
\end{appendix}

\end{document}